\documentclass[12pt]{article}

\usepackage{amssymb}

\usepackage{verbatim}

\def\hH {{\widehat H}_h}

\def\bR {{\mathbb{R}}}
\def\bN {{\mathbb{N}}}

\def\cA {{\mathcal A}}

\def\Di {\displaystyle}
\def\hbr {\hslash}
\def\ccH {{\widehat H}^\hslash}
\def\tH {{\widetilde H}^\hslash}

\makeatletter
\@addtoreset{equation}{section}

\makeatother

\newtheorem{theorem}{Theorem}[section]
\newtheorem{lemma}[theorem]{Lemma}

\newtheorem{remark}[theorem]{Remark}

\begin{document}

\bibliographystyle{plain}

\begin{center}
\Large \bf {
 Born-Oppenheimer-type approximations \\
 for 
  degenerate potentials~:\\
recent results and a survey on the area }
\end{center}

\vskip 0.5cm

 \centerline {\bf {
   Fran\c{c}oise \ TRUC}}

{\it  Universit\'e de Grenoble I, Institut Fourier,\\
            UMR 5582 CNRS-UJF,
            B.P. 74,\\
 38402 St Martin d'H\`eres Cedex, (France), \\
E.Mail: Francoise.Truc@ujf-grenoble.fr }


\begin{abstract}

This paper is devoted to the asymptotics of
eigenvalues for a Schr\"o-dinger operator 
$H_h=-h^2\Delta +~V$ on $L^2({\bf R}^m)$, in the case when the
potential $V$ does not fulfill the non degeneracy condition~:
$V(x) \rightarrow
 +\infty$ as $|x|\rightarrow +\infty$. For such a model, the point
is that the set defined in the phase space by~:~$H_h~\leq~ \lambda$
may  have an infinite volume, so that the Weyl formula which
gives the behaviour of the counting function has to
be revisited.

We recall various results in this area, 
in the classical context ($h=1$ and $\lambda  \rightarrow
+\infty $), as well as in the
semi-classical one ($h~\rightarrow ~0 $) and comment the different methods.    
In section 3, 4 we present our joint works
with A Morame (*),where we consider a degenerate
potential V(x) =f(y) g(z) , where g is assumed to be a homogeneous
  positive function of m variables ,
 smooth outside 0, and f is a smooth  and strictly positive function
 of n variables, with a minimum in 0. 

 In the case where $f(y) \rightarrow
 +\infty$ as $|y|\rightarrow +\infty$, the operator has a compact
 resolvent and we give the asymptotic behaviour, for small values of  h, of the number of eigenvalues less than a fixed energy .  
 
Then, without assumptions on the limit of f, we give a sharp 
estimate of the low eigenvalues, using a Born Oppenheimer
approximation.
 With a refined approach we localize also higher energies . In the
 case when the degree of homogeneity is not less than 2,
we can even assume that the order of these energies is like the inverse
power of the square of h.

Finally we apply the previous methods to a class of
potentials in $R^d$, $d\geq 2$,  which vanish on a regular hypersurface. 

\end{abstract}

\section{Introduction}
  
Let $V$ be a nonnegative, real and continuous potential on ${\bf R}^m$,
 and $h$ a parameter in $]0,1]$. 
The spectral asymptotics of the operator $H_h=-h^2\Delta +V$ on 
$L^2({\bf R}^m)$ have been
intensively studied. More precisely it is well known \cite{reeds} 
that $H_h$ is
essentially selfadjoint with compact resolvent when $V(x) \rightarrow
 +\infty$ as $|x|\rightarrow +\infty$ (we shall say that $V$
 is non degenerate).
Moreover, denoting by $N(\lambda,H_h)$ the number of eigenvalues less
than a fixed energy $\lambda$, the following semiclassical asymptotics
 hold, as $h
 \rightarrow 0$~:
 
\begin{equation} \label{hip}
N(\lambda,H_h)\sim h^{-m}(2\pi)^{-m}v_m\int_{{\bf
    R}^m}(\lambda-V(x))_+^{m/2} dx\ .
\end{equation}
 
 In this formula, $v_m$ denotes the volume of the unit ball in ${\bf
    R}^m$, and the notation $W_+$ means the positive value of $W$.

Let us note that the classical asymptotics are also given by the
formula (\ref{hip}), provided we let $h=1$ and $\lambda  \rightarrow
+\infty $.

In both cases, the result points out the asymptotic correspondance
between the number of eigenstates with energy less than $\lambda$   
and the volume in phase space of the set $\{(x,\xi),f(x,\xi)\leq
\lambda\}$,
where $f(x,\xi)=\xi^2+V(x)$ is the principal symbol of $H_h$.
  
In this paper we propose a review of results concerning the degenerate
case~: the potential $V$ does not tend to infinity with $|x|$, so that
the volume in phase space of the previous set may be infinite.
  
\section{The Tauberian approach}
Let us explain how the problem of the degenerate case came from the
non degenerate one.
 
In 1950 De Wet and Mandl (\cite{de Wet}) proved the formula (\ref{hip})
 in its classical
version, provided $V(x)\geq 1$ and two more conditions on $V$~:

1) a smoothness condition ~:
V differentiable and $|\nabla V|=\circ(V) $

2) a Tauberian type condition~: let $\Phi(V,\lambda)=\int_{{\bf
    R}^m}(\lambda-V(x))_+^{m/2} dx$;
it is assumed that there exists $c$ and $c'$ such that~:

$c\Phi(V,\lambda)\leq\lambda\Phi'(V,\lambda)\leq c'\Phi(V,\lambda)$.

The first condition is local and the second is global. This last
condition was needed to use a Tauberian technique,
which consists on studying  the asymptotic behavior of the Green's
function of the operator $H_1$ and applying a Tauberian theorem.

Refinements were done by Titchmarsh, Levitan and
Kostjucenko,(\cite{tit}, \cite{lev}, \cite{kost}) and
then  Rosenbljum (\cite{rosen}) proved that the formula (\ref{hip}) holds
with ``maximal'' weakening conditions on $V$~:

1) the smoothness condition is replaced by a condition on the
``$L^1$-modulus of continuity'' on unit cubes and by the following assumption~:
$V(y)\leq C' V(x)$ if $|x-y|\leq 1$ .

2) the Tauberian type condition becomes~:
$\sigma(2\lambda,V)\leq C\sigma(\lambda,V)$ (for large $\lambda$),
 where $\sigma(\lambda,V)$ denotes the volume of the set $\{x\in {\bf
    R}^m; V(x)<\lambda \}$.

Solomyak (\cite{sol}) makes the following remark~:

\begin{lemma}\label{hop}

Let $V$ be a positive a-homogeneous potential~: 

$V(x)\geq 0;$  $ V(tx)=t^a V(x)$ for any $ t\geq 0$ ( $a>0 $). 
  
If moreover $V(x)$ is strictly positive ($V(x)\neq 0$ if $ x\neq 0$)
the spectrum of $H_1$ is discrete  and the formula (\ref{hip}) takes
the form~:

$$N(\lambda ,H_1)\sim \gamma_{m,a}\lambda ^{\frac{2m+am}{2a}}\int_{
    S^{m-1}}(V(x))^{-m/a} dx$$
($\gamma_{m,a}$ is a constant depending only on the parameters $m$ and $a$.) 

\end{lemma}
 
From that lemma comes out naturally the idea of investigating the
spectrum whithout the condition
of strict positivity (and thus in a case
of degeneracy of the potential) ; the two main
results 
are (\cite{sol}]~:

\begin{theorem}
 The formula of Lemma \ref{hop} still holds for  a positive
 a-homogeneous potential such that 
$J(V)= \int_{  S^{m-1}}(V(x))^{-m/a} dx$ is finite.

\end{theorem}
The second result deals with a  case where $J(V)$ is infinite~:

let $V(x)=F(y,z)$, $y\in {\bf R}^n$, $z\in {\bf R}^p$, 
$n+p=m$, $m\geq 2$,
such that $F(sy,tz)=s^bt^{a-b}F(y,z)$ (with $0<a<b$) and
$F(y,z)>0$ for $|z||y|\neq 0$.
Denote by $\lambda_j(y)$ the eigenvalues of the operator
$-\Delta_z+F(y,z)$ in $L^2({\bf R}^p)$ and let $s=\frac{2b}{2+a-b}$, 
then~:
\begin{theorem}\label{s}
\begin{eqnarray*}
If\ \frac{n}{b}>\frac{m}{a}\ \ \ N(\lambda ,H_1)&\sim &\gamma_{n,s}\lambda ^{\frac{2m+am}{2b}}\int_{
    S^{m-1}}\Sigma(\lambda_j(y))^{-n/s} dx\\
  if\ \frac{n}{b}=\frac{m}{a}\ \ \
N(\lambda ,H_1)&\sim& \frac{a(a+2)}{2b(a-b)}\gamma_{m,a}
\lambda ^{\frac{2m+am}{2b}}\ln \lambda\int_{
    S^{n-1} S^{p-1}}F(y,z)^{-m/a} dx.
\end{eqnarray*} 
\end{theorem}
 The proof is based on variational techniques and spectral estimates
proved in (\cite{rosen}).But on a heuristic level the result can be
understood in the framework of the theory of Schr\"odinger operators
with operator potential.

This last approach can be found in (\cite{rob}) where D.Robert
extended the theory of
pseudodifferential operators in the form developped by H\"ormander
 to pseudodifferential operators with operator symbols.
It was thus possible to study cases where the operator has a compact
resolvent but the condition $\lim _{\infty}V(x)=+\infty$ is not 
fulfilled. As an example it gives the asymptotics of $N(\lambda ,H_1)$
for the 2-dimensional potential $V(y,z)=y^{2k}(1+z^2)^{l}$, where
$k$ et $l$ are strictly positive. The asymptotics are the following~:

\begin{theorem}
\begin{eqnarray*}
If\, k>l &  N(\lambda ,H_1)\sim \gamma_1
\lambda ^{\frac{l+k+1}{2l}}\\
if\, k=l & N(\lambda ,H_1) \sim \gamma_2
\lambda ^{\frac{2k+1}{2k}}\ln \lambda\\
if\, k<l & N(\lambda ,H_1) \sim \gamma_3
\lambda ^{\frac{2k+1}{2k}}.
\end{eqnarray*}

\end{theorem}

The constants $\gamma_i$ depend only on $k$ and $l$, but the 
first one $\gamma_1$ takes in account the trace of the operator
$(-\Delta_z+z^{2k})^{-(k+1)/2l}$ in $L^2({\bf R})$.

In the  2-dimensional case let us mention the results of B.Simon
(\cite{simon}). He first recalls Weyl's famous result~:
let $H$ be the Dirichlet Laplacian in a bounded region $\Omega$ in
${\bf R}^2$, then the following asymptotics hold~:
$$N(\lambda ,H)\sim  \frac{1}{2}\lambda |\Omega|$$
and then he considers special regions $\Omega$ for which the volume
(denoted by $|\Omega|$) is infinite but the spectrum of the Laplacian
is still discrete.
These regions are of the type~: $\Omega_{\mu}=\{(y,z);
|y||z|^{\mu}\leq 1\}$.

Actually the problem can be derived from the study of the asymptotics
of Schr\"odinger operators with the homogeneous potential~:
$V(y,z) =|y|^{\alpha}|z|^{\beta}.$

In order to get these ``non-Weyl'' asymptotics, he uses
 the Feynman-Kac formula and the Karamata-Tauberian theorem, 
but the main tool is what he calls ``sliced bread
inequalities'', which can be seen as a kind of Born-Oppenheimer
approximation. More precisely  let $H=-\Delta +V(y,z)$ be defined
on ${\bf R}^{n+p}$, and denote by $\lambda_j(y)$ the eigenvalues of
 the operator $-\Delta_z+V(y,z)$ in $L^2({\bf R}^p)$. 
(If the $z$'s are electron coordinates and the $y$'s are nuclear
coordinates, the $\lambda_j(y)$ are the Born- Oppenheimer curves).
 He proves the following lemma~:

$${\rm Tr}e^{-tH}\leq \Sigma_{j} e^{-t(-\Delta_y +\lambda_j(y))}$$  
(when the second term exists).

Thus he gets the two following coupled results~:

\begin{theorem}

If $H=-\Delta +|y|^{\alpha}|z|^{\beta}$ and $\alpha <\beta$, then

$$N(\lambda ,H)\sim c_{\nu}\lambda ^{\frac{2\nu +1}{2}}\ \ \quad 
(\nu=\frac{\beta +2}{2\alpha})$$

{\rm Corollary
}~:
\ \ if $H=-\Delta_{\Omega_{\mu}} $
($\mu >1$), then
$\quad N(\lambda ,H)\sim c_{\mu}\lambda ^{\frac{1}{2\mu +1}}$.
 
\end{theorem}

\begin{theorem}
If $H=-\Delta +|y|^{\alpha}|z|^{\alpha}$, then
$\quad N(\lambda ,H)\sim \frac{1}{\pi}\lambda ^{1+\frac{1}{\alpha}}\ln
\lambda$

{\rm Corollary}~:
 \ \ if $H=-\Delta_{\Omega_{\mu}} $
($\mu =1$), then$\quad N(\lambda ,H)\sim \frac{1}{\pi}\lambda \ln \lambda\ .$
 
\end{theorem}

The constant $c_{\mu}$ depends only on $\mu$, and the constant
$c_{\mu}$ takes in account the trace of the  operator
$(-\Delta_z+|z|^{\beta})^{-\nu}$ in $L^2({\bf R})$.

\section{The min-max approach}

The result presented in this section is based on the method of Courant
and Hilbert, the min-max variational principle. It turns out that this
method can be applied to operators in $L^2({\bf R}^m)$ with principal
symbols which can degenerate on some non bounded manifold of $T^*({\bf
  R}^m)$. It is the case for the Schr\"odinger operator with a
magnetic field $H =(D_x-A(x))^2$, which degenerates on $\{(x,\xi)\in
T^*({\bf R}^m); \xi =A(x)\}$. If the magnetic field $B=dA$ fulfills
the so-called magnetic bottle conditions
(mainly~:~$lim_{\infty}\|B(x)\|=\infty$)
the spectrum is discrete (\cite {avr})
and the classical asymptotics were established by Colin de Verdi\`ere 
(\cite {col}) using the min-max method. The semiclassical version of
the
result is given in (\cite {truc}).

In (\cite {mot}), the min-max method is performed to get semiclassical
asymptotics for a large class of degenerate potentials, namely
potentials of the following form~:
$x=(y,z)\in {\bf R}^n\times {\bf R}^p$, $n+p=m$, $m\geq 2$

 $V(x)=f(y)g(z)$,  $f\in C({\bf R}^n;{\bf R}_+^*)$,
\begin{equation}\label{Hypg}  
g\in C({\bf R}^p;{\bf R}_+), 
\  g(tz)=t^{a}g(z)\ (a>0) \ \forall t> 0\  ,
\ g(z)>0 \ \forall  z\neq 0.
\end{equation}
 The spectrum of the operator $-\Delta_z+g(z)$ in $L^2({\bf R}^p)$  
is discrete and positive. Let us denote by $\mu_j$ its
eigenvalues.
It is easy to make the following remark~:
\begin{remark}\label{o}
If $f(y)\rightarrow +\infty$ as
 $|y|\rightarrow +\infty$ then $H_h=-h^2\Delta +V$ has a compact resolvent.
\end{remark}

Of course if $f$ was supposed to be homogeneous, the asymptotics
would be given by Theorem \ref {s}. Here the assumption on $f$ is only a 
locally uniform regularity~:

$\exists\ b,c>0$ s.t. $c^{-1}\leq f(y)$ and $|f(y)-f(y')|\leq c f(y)
|y-y'|^b$,

for any  $y,y'$ verifying $|y-y'|\leq 1 $.
\begin{theorem}\label{dd}

Let us assume the previous conditions on $f$ and $g$.
 Then there exists $\sigma,\tau \in ]0,1[$ such that, for any
 $\lambda >0$,
one can find $h_0\in ]0,1[$, $C_1,C_2>0$ in order to have
$$(1-h^{\sigma}C_1)n_{h,f}(\lambda-h^{\tau}C_2)\leq N(\lambda;H_h)\leq
(1+h^{\sigma}C_1)n_{h,f}(\lambda+h^{\tau}C_2) \quad \forall h\in ]0,h_0[$$
$$if\\\ n_{h,f}(\lambda)=h^{-n}(2\pi )^{-n}v_n
\int_{{\bf R}^n}\Sigma_{j\in {\bf
    N}}[\lambda-h^{2a/(2+a)}f^{2/(2+a)}(y)\mu_j
]_+^{n/2} dy\ .$$

\end{theorem}

Provided some additional conditions on $f$, the previous result can
be refined as follows ~:

\begin{theorem}\label{de}
 If moreover one can find a constant $C_3$ such that, for any $\mu>1$~:
$$\int_{\{y,f(y)<2\mu\}}f^{-p/a}(y)dy \leq C_3
\int_{\{y,f(y)<\mu\}}f^{-p/a}(y)dy\ ,$$

then one can take $C_2=0$ in Theorem 7:
$$(1-h^{\sigma}C_1)n_{h,f}(\lambda)\leq N(\lambda;H_h)\leq
(1+h^{\sigma}C_1)n_{h,f}(\lambda) \quad \forall h\in ]0,h_0[$$

\end{theorem} 

\begin{remark}
 If moreover $f^{-p/a}\in L^1({\bf R}^n)$ and $g\in C^1({\bf
   R}^p\backslash\{0\})$,
then the formula (1) holds.
\end{remark}

The proof of Theorem \ref {dd} uses a suitable covering of ${\bf R}^n$, so
 that the min-max variational principle allows to deal with Dirichlet
 and Neumann problems in cylinders for the restrained operator
 (with a fixed $y$).
The proof of Theorem \ref {de} is based on an asymptotic formula of the
moment of eigenvalues of $-h^2\Delta_z + g(z)$, which is again obtained
using the min-max principle.
 
As a conclusion, let us notice that if there is some information
on the growth of $f$, then the asymptotics can be computed in
terms of power of $h$:

\begin{remark}
If there exists $k>0$ and $C>0$ such that

$\frac{1}{C}|y|^k\leq f(y) \leq C |y|^k $ for $|y|>1$, then
\begin{eqnarray*}
if\, k>a &  N(\lambda ,H_h)\approx h^{-m}\\
if\, k=a & N(\lambda ,H_h) \approx h^{-m
}\ln \frac{1}{h}\\
if\, k<a & N(\lambda ,H_h) \approx h^{-n-\frac{pa}{k}}
\end{eqnarray*}

\end{remark}

\section{Born-Oppenheimer-type estimates }

In last section we have investigated the asymptotic behavior
of the number of eigenvalues less then
$\lambda $ of $\hH =-h^2\Delta + f(y)g(z).$

Theorem \ref {dd} gives us a hint of what should eigenvalues of $\hH \; $
look like.  This can be done using Born-Oppenheimer-type methods. 

We assume as in last section that :
 $\ g\; \in \; C^\infty ({\bf R}^m\setminus \{ 0\} )\ $
is homogeneous of degree $a>0\; ,$ and 
 assume the following for $f$~:

\begin{equation}\label{Hypf}
\begin{array}{ll}
f\; \in \; C^\infty (\bR ^n),
\ \forall  \alpha \in \bN ^n,\ (|f(y)|+1)^{-1}\partial_y^\alpha f(y)\;
\in L^\infty (\bR ^n)\\
 0\; <\; f(0)\; =\; \inf_{y\in \bR ^n} \; f(y)\\
 f(0)\; <\; \liminf_{|y|\to \infty}\; f(y) \; =\; f(\infty)\\
 \partial ^2f(0)\; >\; 0\\
 \end{array}
 \end{equation}
 $\partial ^2 f(0)$ denotes the hessian matrix in $0$.
\subsection{Using homogeneity}
 By dividing $\hH \ $ by $f(0)\; ,$ we can change the parameter $\ h\ $ and
assume that
 \begin{equation}\label{fzero}
 f(0)=1\; .
 \end{equation}
Let us define~:
$ \hbr \; =\; h^{2/(2+a)}$ and   change $z$ in $z\hbr$;
 we can use the homogeneity of $g$ (\ref{Hypg}) to get~:
 \begin{equation}\label{defHcc}
 sp\; (\hH )\; =\; \hbr^a\;
 sp\; (\ccH)\; ,
 \end{equation}
 with $\Di \ \ccH \; =\; \hbr ^2D_y^2\; +\; D_z^2\; +\; f(y)g(z)\;$.

Let us denote as usually the increasing sequence of eigenvalues
 of $\ D^2_z\; +\; g(z)\; ,$
 (on $L^2(\bR ^m)\; )\; ,$ by $\ (\mu_j)_{j>0 }\; .$\\
 The associated  eigenfunctions will be denoted
 by $\ (\varphi_j)_j\; :$

 By homogeneity (\ref{Hypg}) the eigenvalues
 of $Q_y(z,D_z)=\ D^2_z\; +\; f(y)g(z)\; ,$
 on $L^2(\bR ^m)\; )\; $, for a fixed $y$, are given by the sequence $(\lambda_j(y))_{j>0 }$, where~:
 $\lambda_j(y)=\mu_j\ f^{2/(2+a)}(y) \; .$
 \\
 So as in \cite{MoTr} we get~ :

 \begin{equation}\label{estiH1}
 \ccH\; \geq\; \left [ \;
 \hbr^{2}D^2_y\; +\; \mu_1 f^{2/(2+a)}(y)\; \right ] \; .
 \end{equation}
 This estimate is sharp as we will see below.

Then using  the same kind of estimate as (\ref{estiH1}), one can see that
 \begin{equation}\label{infSpess}
 \inf \; sp_{ess}(\ccH )\; \geq \; \mu_1f^{2/(2+a)}(\infty)\; .
 \end{equation}

 We are in the Born-Oppenheimer approximation situation described
 by A. Martinez in \cite{Ma}~: the "effective "
 potential is given by $\lambda_1(y)=\mu_1\ f^{2/(2+a)}(y)$, the first 
eigenvalue of $Q_y$,
 and the assumptions on $f$ ensure that this potential admits one
 unique and nondegenerate  well  $U=\{ 0\} $,
  with minimal value equal to $\mu_1$. Hence we can apply theorem 4.1 of
\cite{Ma} and get~:

 \begin{theorem}\label{thGround}
 Under the above assumptions, for any arbitrary $C>0$,
 there exists $\ h_0>0 $ such that, if
 $0<\hbr<h_0\; ,$  the operator $(\ccH)$
 admits a finite number of eigenvalues
 $E_{k}(\hbr )$ in $[\mu_1,\mu_1+C\hbr ]$, equal to the number
 of the eigenvalues $e_k$ of
 $\  D_y^2 \; +\;  \frac{\mu_1}{2+a} <~\partial^2f(0)\ y,\ y\ >\ $
   in  $[0,+C]$ such that~:

 \begin{equation}\label{equaGround}
 E_{k}(\hbr )=\lambda_k(\ccH )\; =\; \lambda_k \left ( \hbr^2
 D^2_y+\mu_1f^{2/(2+a)}(y)\right )\; +\; {\bf O}(\hbr^2)
 \; .
 \end{equation}
 More precisely
 $ E_{k}(\hbr )=\ \lambda_k(\ccH)\; $
 has an asymptotic expansion
 \begin{equation}\label{asympHH}
  E_{k}(\hbr )\  \sim \;
 \mu_1\; +\;\hbr\  (\ e_k\ +\  \sum_{j\geq 1}
 \alpha_{kj}\hbr ^{j/2}\;) .
 \end{equation}
 If $ E_{k}(\hbr )\ $ is asymptotically
 non degenerate, then there exists a quasimode
 \begin{equation}\label{quasiHH}
 \phi_k^\hbr (y,z)\; \sim \;
 \hbr ^{-m_k} e^{-\psi (y)/\hbr}
 \sum_{j\geq 0} \hbr^{j/2}a_{kj}(y,z)\; ,
 \end{equation}
 satisfying
 \begin{equation}\label{quasiHHb}
 \begin{array}{ll}
 C^{-1}_{0} \leq \| \hbr^{-m_k}e^{-\psi (y)/\hbr}a_{k0}(y,z)\| \leq C_0\\
 \| \hbr ^{-m_k}e^{-\psi (y)/\hbr}a_{kj}(y,z)\| \leq C_j\\
 \| \left ( \ccH -\mu_1-\;  \hbr e_k - \sum_{1\leq j\leq J}
 \alpha_{kj}\hbr ^{j/2}
 \right ) \\
 \hbr^{-m_k}e^{-\psi (x)/\hbr}
 \sum_{0\leq j\leq J}\hbr ^{j/2}a_{kj}(x,y)\| \; \leq \; C_J \hbr ^{(J+1)/2}
 \end{array}
 \end{equation}
 \end{theorem}
 The formula (\ref{asympHH}) implies
 \begin{equation}\label{asymptGr}
 \lambda_k(\ccH)\; =\; \mu_1\; +\; \hbr \lambda_k
 \left ( D_y^2 + \frac{\mu_1}{2+a}<\ \partial ^2f(0)\ y\ ,\ y\ >\right )
 \; +\; {\bf O}(\hbr ^{3/2})\; ,
 \end{equation}
 and when $k=1\; ,$ one can improve ${\bf O}(\hbr ^{3/2})\; $
 into ${\bf O}(\hbr ^2)\; .$
The function $\psi$ is defined by ~:
$\psi (y)\ =\ d(y, 0)\ $, where $d$ denotes the Agmon distance related
to the degenerate
metric $\mu_1\ f^{2/(2+a)}(y) dy^2.$

\subsection{Improving Born-Oppenheimer methods}

We are interested now with the lower
 energies of $\ccH$ .
 Let us make the change of variables
\begin{equation}\label{changeV}
(y,\; z)\; \to \; (y,\; f^{1/(2+a)}(y)z)\ .
\end{equation}
The Jacobian of this diffeomorphism is $ f^{m/(2+a)}(y)$, so we
perform the change
 of test functions~:
$\Di \; u\; \to \; f^{-m/(4+2a)}(y)u\; ,$ 
  to get a unitary transformation.

Thus we get that
\begin{equation}\label{sptH}
sp\; (\ccH )\; =\; sp\; (\tH )
\end{equation}
where $\tH \ $ is the self-adjoint operator on
$L^2(\bR ^n\times \bR ^m)\ $ given by

\begin{equation}\label{deftH2}
\begin{array}{ll}
\tH \; =\;
\hbr ^2D^2_y\; +\;
f^{2/(2+a)}(y)\left (
D^2_z+g(z)\right )
\\
+\hbr ^2\frac{2}{(2+a)f(y)}(\nabla f(y)D_y)(zD_z)
\\
+i\hbr ^2\frac{1}{(2+a)f^2(y)}\left ( |\nabla f(y)|^2-
f(y)\Delta f(y)\right )[(zD_z)\; -\; i\frac{m}{2}]
\\
+\; \hbr ^2\frac{1}{(2+a)^2f^2(y)}
|\nabla f(y)|^2[(zD_z)^2\; +\; \frac{m^2}{4}]
\end{array}
\end{equation}

  The only significant role up to order 2 in $\hbr$ will be played
actually by the first operator, namely~:
$\tH_1\ =\
\hbr ^2D^2_y\; +\;
f^{2/(2+a)}(y)\left (
D^2_z+g(z)\right )\ $.

This leads to~:

\begin{theorem}\label{thGround2} .

Under the assumptions (\ref{Hypg}) and (\ref{Hypf}),
for any fixed integer $\ N\ >0\ ,$
there exists a positive constant $h_0(N)$ verifying~:
for any $ \hbr \in ]0, h_0(N)[$, for any $k\leq N\; $ and any
$j\leq N\; $  such that
 $$\ \mu_j\; <\; \mu_1f^{2/(2+a)}(\infty)\; ,$$
 there exists an eigenvalue
$\; \lambda_{jk}\; \in \; sp_d \; (\ccH )\ $
 such that
 \begin{equation}\label{equaGround2}
 |\; \lambda_{jk}\; -\; \lambda_k \left ( \hbr^2
   D^2_z+\mu_jf^{2/(2+a)}(z)
\right )\; |\; \leq \; \hbr ^{2}C \; .
 \end{equation}

 Consequently, when $k=1\; ,$ we have
 \begin{equation}\label{equaGround2b}
 |\; \lambda_{j1}\; -\; \left [\mu_j\; +\; \hbr (\mu_j)^{1/2}
 \frac{tr((\partial ^2f(0))^{1/2})}{(2+a)^{1/2}} \right ]\; |\; \leq \;\hbr ^{2}C \; .
 \end{equation}
 \end{theorem}

 \subsection{ Middle energies}

We can refine the preceding results when $a\geq 2$,$\ g\; \in \; C^\infty (\bR ^m)\; $  and $f(\infty)=\infty$.
We  get then sharp localization near the   $\mu_j$'s for much higher
  values of $j$'s. More precisely we prove~:
 \begin{theorem}\label{thGround3} .
Assume the preceding properties, and consider $\ j\ $  such that
$\ \mu_j\; \leq \; \hbr ^{-2}\; ;$ \\
then for any integer $\ N\; ,$ there exists a constant $\ C $
depending only on $\ N\ $ such that, for any
$\ k\; \leq \; N\; ,$ there exists
 an eigenvalue
$\; \lambda_{jk}\; \in \; sp_d \; (\ccH )\ $ verifying
\begin{equation}\label{equaGround3}
|\; \lambda_{jk}\; -\; \lambda_k \left ( \hbr^2 D^2_y+\mu_jf^{2/(2+a)}(y)\right )\; |\; \leq \; C\mu_j \hbr ^{2}\; .
\end{equation}
Consequently, when $k=1\; ,$ we have
\begin{equation}\label{equaGround3b}
|\; \lambda_{j1}\; -\; \left [\mu_j\; +\; \hbr (\mu_j)^{1/2}
\frac{tr((\partial ^2f(0))^{1/2})}{(2+a)^{1/2}} \right ]\; |\; \leq \; C\mu_j \hbr ^{2}\; .
\end{equation}

\end{theorem}

\subsection{An application}

We can apply the previous methods for studying 
  Schr\"odinger operators on $\ L^2(\bR _{s}^{d})\ $
with $\ d\geq 2\; ,$
\begin{equation}\label{defPh}
P^h \; =\; - h ^2 \Delta \; +\; V(s) \;
\end{equation}
with a real and regular potential $\ V(s)\ $ satisfying
\begin{equation}\label{hypV1}
\begin{array}{c}
V\; \in \; C^\infty (\bR ^d \; ;\ [0, +\infty [)\\
\liminf _{|s| \to \infty} \; V(s)\; >\; 0\\
\Gamma \; =\; V^{-1}(\{ 0\} )\quad {\rm is \ a\ regular\
hypersurface.}
\end{array}
\end{equation}

Moreover we assume that $\ \Gamma \ $ is connected and that there exist
$\; m\; \in \; \bN ^\star\ \ {\rm and}\ \ C_0\; >\; 0$ such that
for any $ s$ verifying $ d(s,\ \Gamma )\; <\;
C_{0}^{-1}$ 
\begin{equation}\label{hypV2}
C_{0}^{-1}d^{2m}(s,\ \Gamma ) \; \leq \; V(s)\; \leq \; C_0
\ d^{2m}(s,\ \Gamma )
\end{equation}
$(\ d(E,F)\ $ denotes the euclidian distance between $\ E\ $ and
$\ F\; )\; .$

We choose an orientation on $\ \Gamma \ $ and  a  unit normal
vector $\ N(s)\ $\\
 on each $\ s\; \in \; \Gamma \; ,$
and then,  we can define the function on $\ \Gamma \; ,$
\begin{equation}\label{defF}
f(s)\; =\; \frac{1}{(2m)!}\left ( N(s)\frac{\partial}{\partial
s}\right )^{2m}V(s)\; , \quad \forall \; s\; \in \; \Gamma \; .
\end{equation}
Then by (\ref{hypV1}) and (\ref{hypV2}), $\ f(s)\; >\; 0\; ,\quad
\forall \; s\; \in \; \Gamma \; .$

Finally we assume that the function $\ f\ $ achieves its minimum on
$\ \Gamma \ $ on a finite number of discrete points:
\begin{equation}\label{hypVf1}
\Sigma_0\; =\; f^{-1}(\{ \eta_0\} )\; =\; \{ s_1,\ldots ,\;
s_{\ell_0}\} \; , \quad if \quad \eta_0\; =\; \min_{s\in
\Gamma}\; f(s)\; ,
\end{equation}
and the hessian of $\ f\ $ at each point $\ s_j\; \in \; \Sigma_0\
$ is non degenerate.

$\ Hess(f)_{s_j}\ $ has $\ d-1\ $ non negative eigenvalues
$$  \rho_1^2(s_j)\leq \; \ldots
\; \leq \; \rho_{d-1}^{2}(s_j)\; ,\quad \quad (\; \rho_j(s_j)\;
>\; 0)\; .$$ 
 The eigenvalues
$\ \rho_k^2(s_j)\ $ do not depend on the choice of coordinates. We
denote
\begin{equation}\label{defTrHess}
Tr^+ (Hess(f(s_j)))\; =\; \sum_{\ell =1}^{d-1}\rho_\ell (s_j)\; .
\end{equation}

We denote by $\ (\mu_j)_{j\geq 1}\ $ the increasing sequence of the
eigenvalues of the operator $\Di \ -\; \frac{d^2}{dt^2}\; +\;
t^{2m}\ $
on $\ L^2(\bR )\; .$\\

\begin{theorem}\label{locTh1}
Under the above assumptions, for any $\ N\; \in \; \bN ^\star \;
,$ there exist $\ h_0\; \in \; ]0,1]\ $ and $\ C_0\; >\; 0\ $
such that,
if $\ \mu_j\; <<\; h ^{-4m/(m+1)(2m+3)}\; ,$ \\
and if $\ \alpha \; \in \; \bN ^{d-1}\ $ and $\ |\alpha |\; \leq
\; N\; ,$\\
 then $\ \forall \; s_\ell \;  \in \; \Sigma_0\;
,\quad \exists \; \lambda_{j\ell\alpha }^{h}\; \in \; sp_d(P^h )\ \
\  s.t.$
$$ \left | \;  \lambda_{j\ell\alpha }^{h}\; -\; h ^{2m/(m+1)}
   \left [ \eta^{1/(m+1)}_{0} \mu_j\; +\; h ^{1/(m+1)}\mu_{j}^{1/2}
 \; \cA_\ell (\alpha) \right
] \; \right |$$
$$ \; \leq \; h^2\mu_{j}^{2+3/2m}C_0\; ;$$
with $\displaystyle  \cA_\ell (\alpha) \; =\;
\frac{1}{\eta^{m/(2m+2)}_{0}(m+1)^{1/2}}
\left [ 2\alpha \rho(s_\ell ) \; +\; Tr^+ (Hess(f(s_\ell )))\right ] \; .$
\\
$(\alpha \rho (s_\ell )\; =\; \alpha_1\rho_1(s_\ell )+\ldots
\alpha_{d-1} \rho_{d-1}(s_\ell )\; )\; .$
\end{theorem}

\end{document}